\documentclass[aps,pre,preprint,superscriptaddress]{revtex4-2}
  \usepackage{graphicx}
  \usepackage{bm}
  \usepackage{amsmath,amssymb}
\date{\today}
\title{Repulsive thermal van der Waals interaction in multi-species asymmetric electrolytes driven by external electric fields}
\begin{document}

\title{Repulsive thermal van der Waals interaction in multi-species asymmetric electrolytes driven by external electric fields}
\date{\today}
\author{Guangle Du}
\affiliation{ School of Physical Sciences, University of Chinese Academy of Sciences (UCAS), Beijing 100049, China.}
\author{David S. Dean}\email{david.dean@u-bordeaux.fr}
\affiliation{Universit\'e Bordeaux, CNRS, LOMA, UMR 5798, F-33400 Talence, France.}
\affiliation{ Kavli Institute for Theoretical Sciences, University of Chinese Academy of Sciences (UCAS), Beijing 100049, China.}
\author{Bing Miao}\email{bmiao@ucas.ac.cn}
\affiliation{ Center of Materials Science and Optoelectronics Engineering, College of Materials Science and Opto-Electronic Technology, University of Chinese Academy of Sciences (UCAS), Beijing 100049, China.}
\author{Rudolf Podgornik}
\affiliation{ School of Physical Sciences, University of Chinese Academy of Sciences (UCAS), Beijing 100049, China.}
\affiliation{Wenzhou Institute, University of Chinese Academy of Sciences, Wenzhou, Zhejiang 325001, China.}
\affiliation{Institute of Physics, Chinese Academy of Sciences, Beijing 100190, China.}

\begin{abstract}
It is well established that the long-range component of the thermal van der Waals interaction between two semi-infinite dielectrics becomes short-range when  an  electrolyte is present between them, this is the well known phenomenon of screening.
In [Phys.~Rev.~Lett.~\textbf{133}, 238002 (2024)] it was shown that for a binary symmetric electrolyte, an electric field parallel to the dielectric boundaries  disrupts screening and a long-range thermal repulsive interaction appears. At large applied fields this long-range repulsive interaction can be explained by the fact that the cations and anions have differing average drifts moving in opposite directions, leading to the correlation of charge density fluctuations between the two species to decouple. Here we extend these results to binary electrolytes which are asymmetric as well as electrolytes with more than two ionic species.
\end{abstract}

\maketitle
\section{Introduction}
\label{sec:orga625ece}
Fluctuations, either of quantum or thermal nature, can induce interactions between confining boundaries or intervening objects, known as fluctuation-induced interactions, the most well known examples being the Casimir interaction and its generalization to van der Waals interactions via the Lifshitz theory. These ubiquitous interactions are   fundamental from a theoretical point of view but also have many applications \cite{woodsMaterialsPerspectiveCasimir2016,klimchitskayaCasimirForceReal2009a,woodsPerspectiveRecentFuture2020a}.
van der Waals  interactions are dominant in micro- and nano-electromechanical systems. They may cause device failure by unwanted collapse and adhesion, or can be harnessed to actuate devices
\cite{barcenasScalingMicroNanodevices2005,buksMetastabilityCasimirEffect2001,buksStictionAdhesionEnergy2001a,chanNonlinearMicromechanicalCasimir2001,chanQuantumMechanicalActuation2001,serryRoleCasimirEffect1998}. Tuning Casimir or van der Waals  interactions has thus attracted interest.
From the Lifshitz theory
\cite{lifshitzTheoryMolecularAttractive1956}, it can be shown that van der Waals  interactions can be tuned by changing the optical properties of the dielectrics by using, for example, topological materials with exotic optical properties \cite{woodsPerspectiveRecentFuture2020a,grushinTunableCasimirRepulsion2011}. They can also be tuned by modulating the electrical carrier in semiconductors and graphene \cite{klimchitskayaCONTROLCASIMIRFORCE2011,tseQuantizedCasimirForce2012}.
Another general way to tune, is by forcing the system out of equilibrium  by imposing a temperature difference
\cite{antezzaCasimirLifshitzForceOut2006,antezzaCasimirLifshitzForceOut2008,antezzaNewAsymptoticBehavior2005a,bimonteScatteringApproachCasimir2009,dorofeyevForceAttractionTwo1998,krugerNonequilibriumElectromagneticFluctuations2011,luOutofequilibriumThermalCasimir2015,messinaCasimirLifshitzForceOut2011} between the interacting objects as well as applying external fields \cite{deanNonequilibriumTuningThermal2016,duCorrelationDecouplingCasimir2024}. We also mention that the transient behavior of the Casimir force between two equilibrium regimes at different temperatures induced by temperature  quenches has been studied in electrostatic models and for the critical Casimir effect \cite{gambassiCriticalDynamicsThin2006,gambassiRelaxationPhenomenaCriticality2008,deanNonequilibriumBehaviorPseudoCasimir2009,deanOutofequilibriumBehaviorCasimirtype2010a,deanRelaxationThermalCasimir2014}.

\begin{figure}[!tpb]
\centering
\includegraphics[width=0.9\linewidth]{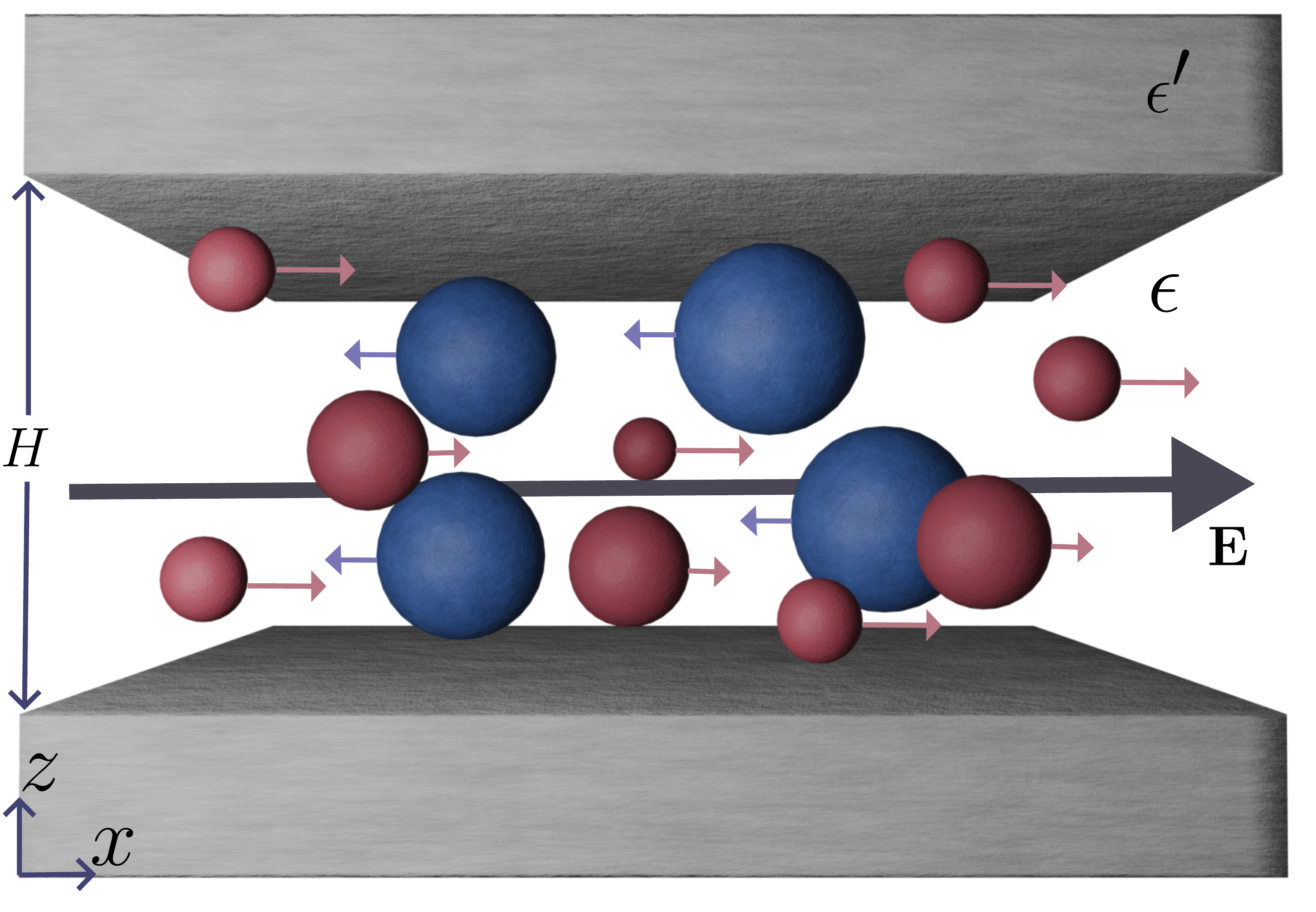}
\caption{\label{fig:schematics}Electrolyte driven by an external electric field \(\mathbf{E}\), in a solvent of dielectric constant \(\epsilon\), confined between two dielectric semi-spaces of dielectric constant \(\epsilon'\). Here \(H\) is the separation between the dielectrics and \(z\) is the normal coordinate with the bottom surface located at \(z=0\). The different sizes of balls and lengths of arrows represent, respectively, the different charges and  bare velocities of different species.}
\end{figure}

When dielectrics are separated by  electrolytes, the thermal component of the van der Waals  interaction is screened and decays exponentially with separation \cite{mahantyDispersionForces1976,parsegianVanWaalsForces2006}. Recently it was shown that strongly applied electric fields parallel to the slab boundaries in an electrolyte can lead to correlation decoupling \cite{duCorrelationDecouplingCasimir2024}, which
can be understood from the fact that  strong external electric fields disrupt the equilibrium  ionic screening atmosphere around the individual ions, reminiscent of the modification of ionic correlations responsible for the first Wien effect studied in strong electrolytes \cite{onsagerWienEffectSimple1957,demeryConductivityStrongElectrolytes2016}.
In \cite{duCorrelationDecouplingCasimir2024} we studied the thermal van der Waals interaction between two parallel dielectric slabs of dielectric constants $\epsilon'$  sandwiching a symmetric binary electrolyte in a solvent of dielectric constant $\epsilon \gg \epsilon'$ and separated by a distance $H$. By a symmetric electrolyte we mean a monovalent electrolyte where the cations and anions have opposite but equal charges but also where the magnitudes of the mobilities of the two ionic species are taken to be identical.
In this paper we generalize the results of \cite{duCorrelationDecouplingCasimir2024} to the systems shown in Fig.~\ref{fig:schematics}. This model for  driven symmetric electrolytes was first proposed by \cite{mahdisoltaniLongRangeFluctuationInducedForces2021} where the van der Waals interaction in the presence of a driving electric field was studied. In  \cite{mahdisoltaniLongRangeFluctuationInducedForces2021} an effective dynamical theory  where the  anisotropy due to the field  is mismatched with the thermal noise was derived. It was established some time ago that  this kind of mismatch  leads to long-range fluctuations and correlations \cite{garridoLongrangeCorrelationsConservative1990,grinsteinConservationLawsAnisotropy1990}, and it was subsequently shown that long-range fluctuation  induced interactions can be found in such systems \cite{kirkpatrickGiantCasimirEffect2013,kirkpatrickFluctuationinducedPressuresFluids2014}.

We consider an electrolyte consisting of \(N\) species of ions with different densities \(\rho_i(\mathbf{x})\), charges \(q_i\), and diffusion constants \(D_i\) with \(i\in \{1,2,\ldots,N\}\) labeling different species.
The electrolyte is in solvent of dielectric constant \(\epsilon\) and is confined between two dielectric semi-spaces (slabs) of dielectric constant \(\epsilon'\). In the following, we will consider the common limit  for aqueous electrolytes \(\epsilon \gg \epsilon'\) which gives rise to  the Neumann boundary conditions for densities and the corresponding electric potential at the dielectric surfaces \cite{duCorrelationDecouplingCasimir2024}. Crucially this assumption allows us to perform a Fourier series expansion in the normal direction.
An external electric field \(\mathbf{E}\) is applied parallel to dielectric boundaries, here, along \(x\)-direction.
The ions are deemed to perform Brownian motion with electrostatic interactions among them due to the fields they create themselves as well as the imposed external field. The choice of Brownian  dynamics allows us to analyze the model using  stochastic density functional theory \cite{kawasakiStochasticModelSlow1994,deanLangevinEquationDensity1996}. We will furthermore assume that there is no surface charge present in the system. This assumption allows an expansion of the  stochastic density functional theory about the mean density $\overline \rho_i$ of each species. 

In the precise dielectric limit mentioned above and in the absence of electrolyte, the thermal component of the van der Waals force per unit area is attractive and universal and given by
\begin{equation}
f_{\rm vdW}(H)= -\frac{k_{\mathrm{B}}T\zeta(3)}{8\pi H^3},\label{vdW}
\end{equation}
where $\zeta(z)$ denotes the Riemann zeta function.
When an electrolyte is added, the thermal component is screened \cite{mahantyDispersionForces1976,parsegianVanWaalsForces2006}  and is given by
\begin{equation}
f_{\rm sc}(H)=-\frac{k_BT }{4\pi H^3} \int_{\kappa H}^{\infty} \mathrm{d} K \,K^2\left(\coth K - 1 \right),
\end{equation}
 and $\kappa$ is the inverse Debye screening length given by
\begin{equation}
\kappa^2 =  \sum_{i=1}^N\frac{\beta q_i^2 \bar{\rho}_i}{\epsilon},
\end{equation}
where $\overline \rho_i$ is the mean density of the ionic species denoted by $i$, $\beta = 1/k_B T$ and recall that $\epsilon$ is the dielectric constant of the solvent. The overall condition of electroneutrality imposes the constraint
\begin{equation}
\sum_{i}\overline \rho_i q_i =0.
\end{equation}
In the limit where $H\kappa\gg1$ (so the screening length is much smaller than the separation of the dielectrics), one can explicitly see the exponential screening in the leading asymptotic term
\begin{equation}
f_{\rm sc}(H)=-\frac{k_BT \kappa^2}{4\pi  H} \exp(-2\kappa H).
\end{equation}

There are many  derivations of the equilibrium screened  van der Waals  interaction for parallel dielectric slabs separated by electrolyte solutions using the Lifshitz theory or field theoretical methods \cite{mahantyDispersionForces1976,parsegianVanWaalsForces2006,netzStaticVanWaals2001,jancoviciScreeningClassicalCasimir2004,deanElectrostaticFluctuationsSoap2002}.  In equilibrium these methods essentially treat the charge densities due to dipoles and charge density due to ions on the same footing, and are typically written in terms of the fluctuating electrostatic potential. Another approach, which for completeness we will expound in the following section, is to integrate out the dipole degrees of freedom to write down an effective theory for the ionic degrees of freedom. This approach is not necessary in an equilibrium system, but this decomposition is essential to treat the non-equilibrium case where an electric field is applied. In equilibrium however, this approach is still rather striking from a physical point of view. Using the dipole/ion decomposition we will show that the total thermal van der Waals interaction is given by a purely van der Waals interaction $f_{\rm vdW}$ given by Eq.~(\ref{vdW}) (that would be there in the absence of an intervening electrolyte), plus a contribution coming from the ionic charge density fluctuations denoted by $f_{\rm ion}$. Note that the effective theory describing the ions incorporates the effect of the dielectric discontinuity via the fact that the Coulomb interaction is modified by the formation of image charges in the surrounding dielectrics. The total force per unit area is thus $f_{\rm t} = f_{\rm vdW} + f_{\rm ion}$, and remarkably one finds that
\begin{equation}
f_{\rm ion}= f_{\rm sc} - f_{\rm vdW},
\end{equation}
and so we recover the screened result $f_{\rm t}= f_{\text{sc}}$. Therefore, in terms of the decomposition of the forces due to dipole and density fluctuations, screening is achieved by an {\em exact} cancellation between an attractive long-range interaction due to dipole fluctuations and a repulsive long-range interaction due to the ionic charge fluctuations. This screening means that  the thermal contribution to the van der Waals force is drastically reduced at large separations $H$, but we see that it is due  to a rather miraculous cancellation mechanism. 

When an electric field is applied parallel to the plates, an electric current flows, and cations and anions move in opposite directions, and the system is clearly out of equilibrium. If we denote by $D_i$ the diffusion constant of the ionic species $i$, then when the applied field $E$ is large it dominates over the field generated by the other ions and the {\em bare} velocity of ionic species $i$, parallel to the dielectric slabs, is then given  by
\begin{equation}
v_i = \beta D_i q_i E,
\end{equation}
the term $\mu_i = \beta D_i$ corresponding to the mobility of the ionic species $i$.

For symmetric electrolytes  ($q_{1}=-q_{2}=q$, $D_{1}=D_{2}=D$, $\overline \rho_1= \overline \rho_2=\overline \rho$) it was shown that the applied electric field induces a long-range repulsive interaction~\cite{duCorrelationDecouplingCasimir2024}
\begin{equation}
f^{\text{lr}}_{\text{t}}(H) =
\frac{k_B T\zeta(3)}{8\pi  H^3} \left[ 1 - \frac{1}{\sqrt{1+ \frac{\beta \epsilon E^2}{2\bar{\rho}}}} \right].
\end{equation}
Furthermore in the limit $E\to\infty$ we see that 
\begin{equation}
f^{\text{lr}}_{\text{t}}(H, E\to\infty) =
\frac{k_B T\zeta(3)}{8\pi  H^3}.
\end{equation}
In \cite{duCorrelationDecouplingCasimir2024} we explained this result physically from an equilibrium type argument based on the idea that charges which move with different bare velocities decouple or effectively no longer interact when they have very different bare velocities, which is clearly the case at very large applied fields. The argument goes as follows, as charges with the same bare velocity are moving along together they are able to interact and equilibrate and screen amongst themselves. Charges with different bare velocities do not equilibrate with each other and effectively do not interact. In mathematical terms the correlation between charges with different bare velocities is zero so they are essentially independent. Each type of charge density fluctuation is effectively in equilibrium with itself as it is simply in a Galilean frame with the constant bare velocity with respect  to the laboratory frame. Note that one could have accidental degeneracies (though this would require remarkable fine tuning)  among different charges, $v_i=v_j$ for  $i$ and $j$ say when $\beta D_i q_i= \beta D_j q_j$, in this case the two different charge types equilibrate together and
give a single repulsive van der Waals interaction.
Each independent charge density fluctuation thus gives a repulsive van der Waals contribution at large applied fields. When there are two independent charge density fluctuations, thus taking off the attractive van der Waals contribution and leads to a net single repulsive van der Waals interaction. This physical argument then implies that if one had $N$ ionic species with different bare velocities at large applied fields one would expect
\begin{equation}
f^{\text{lr}}_{\text{t}}(H, E\to\infty) =
(N-1)\frac{k_B T\zeta(3)}{8\pi  H^3}.\label{n-1}
\end{equation}

In this paper, we investigate the van der Waals interactions in multi-species electrolytes driven by external electric fields.
We find that in the strong external electric field limit, the correlation between different species with different bare velocities  are decoupled and those with the same bare velocity are still correlated as in the equilibrium case. Consequently, each decoupled species contributes a repulsive force
with the magnitude identical to that of van der Waals interaction and those species still coupled as a whole contributes a repulsive force with one magnitude of van der Waals interaction, this means that Eq.~(\ref{n-1}) does indeed hold.

The paper is organized as follows.
In section \ref{partion_function}, we show how the van der Waals interaction for electrolyte systems can be derived by decomposing it into pure van der Waals or dipolar contributions along  with the ionic contribution. This is done at the level of Debye-H\"uckel theory which essentially is an expansion of weak fluctuations about the mean density. The treatment recovers the results of  \cite{jancoviciScreeningClassicalCasimir2004} where the decomposition into dipolar and ionic contributions was explicitly made, it also generalizes their results to multi-component electrolytes.
The decomposition is also of use for the dynamical treatment in section \ref{SDFT} as it provides us with a theory for the ionic dynamics assuming that the dielectric or dipole components of the problem are instantaneously in equilibrium with any given distribution of the ionic charges. 
In section \ref{SDFT}, stress tensor dynamics of the density field fluctuations are derived from stochastic density functional theory and expanded, as in the static case, in terms of weak density fluctuations yielding an analytically soluble theory. We also derive the stress tensor needed to compute the resulting van der Waals interaction for the driven system.
 The resulting, average, van der Waals  interaction is then expressed in terms of the  density correlations in the steady state, and these correlation functions are computed from a  Lyapunov equation we derive.
In section \ref{equilibrium}, the equilibrium screened interaction for multi-species electrolytes is reproduced from the dynamical method.
In section \ref{large-E}, the Lyapunov equation is solved in large \(E\) limit by perturbation expansion and van der Waals  interaction is calculated for both non-degenerate and degenerate cases depending on the bare velocities. In section \ref{general-E}, the Lyapunov equations for binary and ternary electrolytes are explicitly solved and the corresponding long-range van der Waals interactions for external electric field of general strength are given. The main results are summed up in section \ref{conclusions}.
\section{Partition function for equilibrium system without external electric fields}
\label{partion_function}
We start from the partition function for the equilibrium system without external electric fields \cite{jancoviciScreeningClassicalCasimir2004,duCorrelationDecouplingCasimir2024}
\begin{align}
Z = &\, \int  \mathcal{D}[\phi]\, \prod_{i=1}^{N} \mathcal{D}[\rho_i]\exp \Bigg\{ - \sum_{i=1}^N \int \mathrm{d} \mathbf{x} \,\rho_i(\mathbf{x}) \left[ \ln \rho_i(\mathbf{x}) -1 \right] \nonumber\\
&\, - \beta \int \mathrm{d}\mathbf{x} \left[ \frac{1}{2} \epsilon(\mathbf{x}) (\nabla \phi(\mathbf{x}))^2 - \mathrm{i} \phi(\mathbf{x}) \rho_{\text{c}}(\mathbf{x}) \right] \Bigg\},
\end{align}
where \(\epsilon(\mathbf{x})\) is the dielectric function, \(\phi(\mathbf{x})\) is the electric potential field, and \(\rho_{\text{c}}(\mathbf{x}) = \sum_i q_i \rho_i(\mathbf{x})\) is the charge density.
Carrying out the Gaussian integral over \(\phi(\mathbf{x})\), we can transform the partition function into the product of field and ionic parts
\(Z = Z_{\text{vdW}} Z_{\text{ion}}\),
where
\begin{align} \label{eq:Z-vdW}
Z_{\text{vdW}} = \int \mathcal{D}[\phi] \exp \left\{ - \frac{1}{2} \beta \int \mathrm{d}\mathbf{x} \epsilon(\mathbf{x}) \left[ \nabla \phi(\mathbf{x}) \right]^2 \right\}
\end{align}
is the partition function for the van der Waals interaction in the absence of any ionic charge density
and
\begin{align} \label{eq:Z-ion}
Z_{\text{ion}} = \int \prod_{i=1}^N \mathcal{D}[\rho_i]\exp \left\{-\beta H_{\text{ion}}[\rho_i]  \right\},
\end{align}
is the partition function for the ions with \(H_{\text{ion}}[\rho_i]\) being the ionic Hamiltonian
\begin{align}
H_{\text{ion}}[\rho_i]
=&\, k_{\text{B}}T \sum_{i=1}^N \int \mathrm{d} \mathbf{x} \,\rho_i(\mathbf{x}) \left[ \ln \rho_i(\mathbf{x}) -1 \right] \nonumber\\
&\, + \frac{1}{2} \int \mathrm{d}\mathbf{x} \mathrm{d}\mathbf{x}' \rho_{\text{c}}(\mathbf{x}) G(\mathbf{x},\mathbf{x}') \rho_{\text{c}}(\mathbf{x}').
\end{align}
\(G(\mathbf{x},\mathbf{x}')\) is the Green's function satisfying
\begin{align}
\nabla\cdot \epsilon(\mathbf{x}) \nabla G(\mathbf{x},\mathbf{x}') = - \delta(\mathbf{x}-\mathbf{x}').\label{gf}
\end{align}
In what follows we will assume that the dipole degrees of freedom are instantaneously in equilibrium with any fixed ionic charge distribution. This is reasonable as the dipolar degrees of freedom are electronic and the ionic degrees of freedom are governed by Brownian motion which will be much slower. Furthermore,  this assumption has  two important physical consequences, first it means that there is always a thermal van der Waals contribution due to the dipoles in the dielectric media which can be derived from the free energy corresponding to the partition function  $Z_{\rm vdW}$ in Eq.~(\ref{eq:Z-vdW}). Secondly, the effective interaction between the ions is not the  pure Coulomb potential but that given from the Green's function in Eq.~(\ref{gf}) which includes the interaction with the image charges in the surrounding dielectrics.
 
We now expand the ionic densities around their mean values \(\bar{\rho}_i\), which are independent of positions as we assume that there is no surface charge in the system,
\begin{align}
\label{eq:fluctuations-expansion}
\rho_i(\mathbf{x}) = \bar{\rho}_i + n_i(\mathbf{x}).
\end{align}
The ionic Hamiltonian to the second order in \(n_i\)
is
\begin{align}
Z_{\text{ion}} = &\, \exp \left[ - V \sum_{i=1}^N \bar{\rho}_i (\ln \bar{\rho}_i -1)  \right] \nonumber\\
 &\, \times \int \mathcal{D}[n_i] \exp \left\{-\beta H_{\text{ion}}[n_i] \right\},
\end{align}
where the first term corresponds to the partition function of \(N\) ideal gases of density \(\bar{\rho}_i\) in the total volume \(V\) occupied by the ions
and \(H_{\text{ion}}[n_i]\) is the ionic Hamiltonian for density fluctuations
\begin{align}
H_{\text{ion}}[n_i] = &\, \frac{k_{\text{B}}T}{2} \int \mathrm{d}\mathbf{x} \sum_{i=1}^N \frac{[n_i(\mathbf{x})]^2}{\bar{\rho}_i} \nonumber\\
&\, + \frac{1}{2} \int \mathrm{d}\mathbf{x}\mathrm{d}\mathbf{x}'  \rho_{\text{c}}(\mathbf{x})G(\mathbf{x},\mathbf{x}') \rho_{\text{c}}(\mathbf{x}').
\end{align}
Here the charge density reduces to \(\rho_{\text{c}}(\mathbf{x}) = \sum_{i=1}^N q_i n_i(\mathbf{x})\) due to the electroneutrality  condition \(\sum_{i=1}^N q_{i} \bar{\rho}_i =0\), thus giving
\begin{align}
&H_{\text{ion}}[n_i] = \, \frac{k_{\text{B}}T}{2} \int \mathrm{d}\mathbf{x} \sum_{i=1}^N \frac{[n_i(\mathbf{x})]^2}{\bar{\rho}_i} \nonumber\\
&\, + \frac{1}{2} \int \mathrm{d}\mathbf{x}\mathrm{d}\mathbf{x}' \left[\sum_{i=1}^N q_i n_i(\mathbf{x})\right]G(\mathbf{x},\mathbf{x}') \left[\sum_{i=1}^N q_i n_i(\mathbf{x}')\right].
\end{align}
The above is thus the effective Hamiltonian for the density fluctuations.
\section{Stochastic density functional theory}
\label{SDFT}
When an external electric field is applied, the system is driven out of equilibrium and one must use a dynamical formulation of the problem. Here we will assume that the ions obey Brownian dynamics in an aqueous solvent, the forces acting on the ions are due to the applied electric field and the electrostatic interactions between the ions.
We resort to a stochastic density functional theory where the dynamics of ion densities is governed by \cite{deanLangevinEquationDensity1996}
\begin{align}
\label{eqs:DK-nonlinear}
\partial_t \rho_i(\mathbf{x},t) = \nabla \cdot \Bigg[ & \beta D_i \rho_i \nabla \frac{\delta  H_{\text{ion}}[\rho_i]}{\delta \rho_i} - \beta D_i q_i \rho_i \mathbf{E} \nonumber\\
& + \sqrt{2D_i \rho_i} \, \bm{\eta}_i(\mathbf{x},t)\Bigg],
\end{align}
where \(\bm{\eta}_i(\mathbf{x},t)\) are spatio-temporal Gaussian white noise vector fields
\begin{align}
\langle \eta_{i,\alpha}(\mathbf{x},t) \eta_{j,\beta}(\mathbf{x}',t')\rangle = \delta_{ij} \delta_{\alpha\beta} \delta(\mathbf{x}-\mathbf{x}') \delta(t-t'),
\end{align}
with \(i\) and \(j\) being the species indices, \(\alpha\) and \(\beta\) being the vector element indices, respectively.
The presence of the dielectric boundaries in the problem means that Eqs.~(\ref{eqs:DK-nonlinear}) must be supplemented with the no flux boundary conditions for  deterministic part of the current
\begin{equation}
\beta D_i \rho_i \partial_z \frac{\delta  H_{\text{ion}}[\rho_i]}{\delta \rho_i}=0, \label{f1}
\end{equation}
and the random part of the current
\begin{equation}
{\eta}_{iz}(\mathbf{x},t)=0, \label{f2}
\end{equation}
at each dielectric interface. Note that the noise part of the current is independent of the
deterministic current as Eqs.~(\ref{eqs:DK-nonlinear}) are to be understood in the Ito interpretation of stochastic calculus, meaning that both currents must have no flux boundary conditions and not just the total sum.
One can read off from the above dynamical equations the total body force on all species of ions as the divergence of total stress tensor \(\bm{\sigma}_{\text{t}}\) \cite{krugerStressesNonequilibriumFluids2018}
\begin{align}
\nabla \cdot \bm{\sigma}_{\text{t}} = - \sum_{i=1}^N \rho_i \nabla \frac{\delta  H_{\text{ion}}[\rho_i]}{\delta \rho_i}.
\end{align}
We restrict our attention to the part of the  stress tensor \(\bm{\sigma}\) generated by  density fluctuations and neglect $H$-independent bulk terms such as the ideal gas pressure part, this leads to 
\begin{align}
\nabla\cdot \bm{\sigma} &= -  \sum_{i=1}^N n_i \nabla \frac{\delta H_{\text{ion}}}{\delta n_i} \nonumber\\
&= - \sum_{i=1}^N \frac{n_i \nabla n_i}{\beta\bar{\rho}_i}  - \sum_{i=1}^N q_i n_i \nabla \phi,
\end{align}
where $\phi$ is the electrostatic potential generated by the ionic charges
\begin{align}
\phi(\mathbf{x})=\int d\mathbf{x}'G(\mathbf{x},\mathbf{x}')\rho_c(\mathbf{x}').
\end{align}
Using straightforward algebra  one can immediately extract
\begin{align} \label{eq:stress-ion}
\sigma_{ij} = - \delta_{ij}\sum_{k=1}^N \frac{n_k^2}{2\beta\bar{\rho}_k}  + \epsilon \left[ \nabla_i \phi \, \nabla_j \phi - \frac{1}{2} \delta_{ij} \left( \nabla \phi \right)^2 \right],
\end{align}
where the first term is van't Hoff osmotic stress from the density fluctuations and the second term is just the Maxwell stress tensor.
In the limit \(\epsilon \gg \epsilon'\), the boundary condition for \(\phi\) is Neumann, \emph{i.e.}, \(\partial_z \phi|_{z=0} = 0\). Then the normal force on the boundary due to ion density fluctuations is
\begin{align} \label{eq:force-ion}
f_{\text{ion}} = - \langle \sigma_{zz}\rangle|_{z=0} = \left.
\left\langle\sum_{i=1}^N \frac{ n_i^2}{2\beta\bar{\rho}_i}  + \frac{\epsilon}{2}
 \left( \nabla_{\|} \phi \right)^2 \right\rangle\right|_{z=0},
 \end{align}
where \(\nabla_{\|}\) denotes the gradient in the \(x\)-\(y\) plane.

Eqs.~\eqref{eqs:DK-nonlinear} are nonlinear and hard to solve, but become tractable after linearization
via expanding the ion densities as in Eq.~\eqref{eq:fluctuations-expansion}
\begin{align}
\partial_t n_i =&\, D_i \left( \nabla^2 n_i - \frac{\beta q_i \bar{\rho}_i}{\epsilon} \sum_j q_j n_j - \beta q_i \mathbf{E} \cdot \nabla n_i \right) \nonumber\\
& + \nabla \cdot \sqrt{2D_i \bar{\rho}_i} \, \bm{\eta}_i.
\end{align}
It is easy to see from the full no flux boundary condition in Eq. (\ref{f1})  that  boundary conditions for \(n_i\) are  Neumann, because, crucially, our choice of dielectrics means that those for  \(\phi\) are also Neumann.
This means that we can use  the following Fourier expansion
\begin{align}
n_i(\mathbf{x}_{\|}, z, t) =&\, \int \frac{\mathrm{d}\mathbf{k}}{(2\pi)^2} \sum_{n=0}^{\infty} \frac{1}{\sqrt{N_n}} \tilde{n}_{i,n}(\mathbf{k}, t) \nonumber\\
& \times \exp(\mathrm{i} \mathbf{k}\cdot \mathbf{x}_{\|}) \cos (p_n z),
\end{align}
where \(N_0 = H\) and \(N_n = H/2\) for \(n\ge 1\), \(p_n = n\pi/H\) enforces the Neumann boundary conditions. In Fourier space, we obtain
\begin{align}
\partial_t \tilde{n}_{i,n} =&\, - D_i(k^2 + p_n^2 + \beta q_i E \mathrm{i} k_x) \tilde{n}_{i,n} \nonumber\\
&\, - D_i \frac{\beta q_i \bar{\rho}_i}{\epsilon} \sum_j q_j \tilde{n}_{j,n} + \xi_{i,n}(\mathbf{k},t).
\end{align}
In the above, the noise term has correlation function
\begin{equation}\langle \xi_{i,n}(\mathbf{k}, t) \xi_{j,m}(\mathbf{k}', t')\rangle = (2\pi)^2 \delta(t-t') \delta(\mathbf{k}+\mathbf{k}') \delta_{ij} \delta_{nm} 2 D_i \bar{\rho}_i (k^2 + p_n^2).
\end{equation}
 
We denote  the equal-time correlation functions in the steady state as
\begin{align}
\langle \tilde{n}_{i,n} (\mathbf{k}) \tilde{n}_{j,n}(\mathbf{k}')\rangle = (2\pi)^2 \delta(\mathbf{k}+\mathbf{k}') \tilde{C}_{ij,n}(\mathbf{k}),
\end{align}
where the correlation matrix \(\tilde{C}_{ij,n}(\mathbf{k})\) obeys the Lyapunov equation \cite{zwanzigNonequilibriumStatisticalMechanics2001}
\begin{align}
\label{eq:Lyapunov}
A \tilde{C}_n + \tilde{C}_n A^{\dagger} = 2R,
\end{align}
with
\begin{subequations}
\begin{align}
A_{ij} &= D_i \left[  (k^2 + p_n^2 + \beta q_i E \mathrm{i} k_x) \delta_{ij} + \frac{\beta q_i \bar{\rho}_i}{\epsilon} q_j \right], \\
R_{ij} &= (k^2 + p_n^2) D_i \bar{\rho}_i \delta_{ij}.
\end{align}
\end{subequations}
In terms of the correlation functions, the  van der Waals force due to ion fluctuations given by Eq. (\ref{eq:force-ion}) can then be written as
\begin{align}
f_{\text{ion}} = \frac{2}{H} k_{\text{B}} T \int \frac{\mathrm{d}\mathbf{k}}{(2\pi)^2} \sideset{}{'}\sum_{n=0}^{\infty} S_n(\mathbf{k}),
\end{align}
where
\begin{align}
S_n = \sum_i \frac{\tilde{C}_{ii,n} }{2\bar{\rho}_i} + \frac{\beta k^2}{2 \epsilon (k^2 + p_n^2)^2} \sum_{i,j} q_i q_j \tilde{C}_{ij,n},  \label{snde}
\end{align}
and the prime indicates that there is an extra \(1/2\) factor for \(n=0\).

To facilitate the calculation of \(S_n\), we derive a relation between the van't Hoff osmotic and the Maxwell contributions to the stress, corresponding to the first and second term in Eq.~(\ref{snde}), respectively. Looking at the diagonal elements on both sides of Eq.~\eqref{eq:Lyapunov}, we obtain
\begin{align}
\frac{\tilde{C}_{ii}}{\bar{\rho}_i} + \frac{\beta}{\epsilon(k^2 + p_n^2)} \sum_j q_i q_j \frac{1}{2} (\tilde{C}_{ji} + \tilde{C}_{ij}) = 1,
\end{align}
where we have omitted the Fourier series index to avoid cluttering notations.
Note that the term with \(E\) vanishes.
Summation over \(i\) then leads to
\begin{align}
\sum_i \frac{\tilde{C}_{ii}}{\bar{\rho}_i} + \frac{\beta}{\epsilon(k^2 + p_n^2)} \sum_{ij} q_i q_j \tilde{C}_{ij} = N,
\end{align}
a general identity which holds even in the presence of the  external electric field.
Using this in Eq. (\ref{snde}), \(S_n\) can be expressed solely in terms of  diagonal elements of the correlation matrix
\begin{align} \label{eq:Sn-alternative}
S_n = \frac{1}{2} \left( N \frac{k^2}{k^2 + p_n^2} + \frac{p_n^2}{k^2 + p_n^2} \sum_i \frac{\tilde{C}_{ii}}{\bar{\rho}_i} \right).
\end{align}
\section{Equilibrium van der Waals interaction}
\label{equilibrium}
Using the stochastic density functional theory expounded in the last section, we first of all reproduce the van der Waals interaction in the equilibrium case with \(E=0\), where
the elements of matrix \(A\) reduce to
\(A_{ij} = D_i \left[  (k^2 + p_n^2 ) \delta_{ij} + \frac{\beta q_i \bar{\rho}_i}{\epsilon} q_j \right]\).
It is straightforward to show that \(A R = R A^{\text{T}}\) holds, so Eq.~\eqref{eq:Lyapunov} has the solution
\(\tilde{C} = A^{-1}R\), written in element form as
\begin{align}
\label{eq:C0-reduced}
\sum_j \left(\delta_{ij} + \frac{1}{k^2+p_n^2} \frac{\beta q_i \bar{\rho}_i}{\epsilon} q_j   \right) \tilde{C}_{jk} = \bar{\rho}_i \delta_{ik}.
\end{align}
Note that this can also be directly derived from the equilibrium ion partition function \(Z_{\text{ion}}\) in Eq.~\eqref{eq:Z-ion}.
We introduce
\(B_{ij} = \frac{1}{k^2+p_n^2} \frac{\beta q_i \bar{\rho}_i}{\epsilon} q_j\),
note that since \(B\) is of an outer product form, it has only one non-zero eigenvalue
\(\kappa^2/(k^2 + p_n^2)\) with \(\kappa^2 \equiv \sum_{i=1}^N\beta q_i^2 \bar{\rho}_i/\epsilon\) being the inverse square of Debye length.
Therefore,
\begin{align}
\tilde{C}_{ij} =
 (I+B)^{-1}_{ij} \bar{\rho}_j .
\end{align}
Setting \(j=i\) and summing over \(i\) leads to
\begin{align}
\sum_i \frac{\tilde{C}_{ii}}{\bar{\rho}_i}=
\text{Tr} \left[ (1+B)^{-1} \right]
= N - \frac{\kappa^2}{k^2 + p_n^2 + \kappa^2}.
\end{align}
Substituting this into Eq.~\eqref{eq:Sn-alternative}, we obtain
\begin{align}
S_n = \frac{1}{2} \left( \frac{k^2}{k^2 + p_n^2} - \frac{k^2 + \kappa^2}{k^2 + p_n^2 + \kappa^2}   \right),
\end{align}
where we have discarded bulk terms that are irrelevant for the $H$-dependent van der Waals interaction.
The van der Waals contribution to the stress corresponding to the partition function \(Z_{\text{vdW}}\) in Eq.~\eqref{eq:Z-vdW} is
\begin{align}
f_{\text{vdW}} = \frac{2}{H} k_{\text{B}} T \int \frac{\mathrm{d}\mathbf{k}}{(2\pi)^2} \sideset{}{'}\sum_{n=0}^{\infty} \left[ -\frac{1}{2} \frac{k^2}{k^2 + p_n^2} \right].
\end{align}
The sums appearing in the expressions above can be evaluated using the identity
\begin{equation}
\sideset{}{'}\sum_{n=0}^\infty \frac{k^2}{p_n^2 + k^2}=\frac{1}{2}kH\coth(k H).
\end{equation}
Now taking the  sum of the ionic and van der Waals contributions  and subtracting a remaining  bulk term  \cite{lifshitzStatisticalPhysics1980}, we obtain
the total van der Waals interaction
\begin{align}
f_{\text{t}} = - \frac{k_{\text{B}}T}{4\pi H^3} \int_{H\kappa}^{\infty} \mathrm{d} K K^2 (\coth K -1),
\end{align}
which is nothing but the equilibrium screened attractive force.
\section{Correlation decoupling due to strong external electric fields}
\label{large-E}
It has been shown that in a binary symmetric electrolyte, the correlation decoupling due to strong external electric fields leads to a long-range repulsive van der Waals interaction with a magnitude the same as that of van der Waals interaction without any electrolyte \cite{duCorrelationDecouplingCasimir2024}.
There, the correlations were first worked out by solving the corresponding Lyapunov equation and then the large \(E\) limit was taken.
However, for multi-species electrolytes with an arbitrarily large number of ion species, solving the corresponding nonequilibrium version of Lyapunov equation seems beyond reach.
Here, we instead take the limit of large \(E\) from the beginning by performing perturbation expansion.
We will defer the discussion of the specific binary and ternary asymmetric species cases in section \ref{general-E}.

We first split coefficient matrix \(A\) into the following two parts
\begin{align}
A = A_1 + E A_2
\end{align}
with
\(A_{1,ij} = D_i \left[  (k^2 + p_n^2 ) \delta_{ij} + \frac{\beta q_i \bar{\rho}_i}{\epsilon} q_j \right]\),
and
\(A_{2,ij} = D_i \beta q_i \mathrm{i} k_x \delta_{ij}\).
When \(E\) is large, we can perform perturbation expansion with \(\varepsilon \equiv 1/ E\) being the small parameter
\begin{align}
\tilde{C} = \sum_{i=0}^\infty \tilde{C}^{(i)} \varepsilon^i.
\end{align}
The Lyapunov equation becomes
\begin{align}
(\varepsilon A_1 + A_2) \tilde{C} + \tilde{C} (\varepsilon A_1^{\text{T}} - A_2) = 2 \varepsilon R,
\end{align}
where we have used the facts that \(A_1^\dagger = A_1^{\text{T}}\) and \(A_2^\dagger = -A_2\).
Then at zeroth order of \(\varepsilon\), we have
\(A_2 \tilde{C}^{(0)} - \tilde{C}^{(0)} A_2 = 0\),
written in element form as
\begin{align} \label{eq:zero-order}
(A_{2,ii} - A_{2,jj}) \tilde{C}_{ij}^{(0)} = 0,
\end{align}
with no summation implied here over repeated indices.
The following derivations depend on if there exists \(i\neq j\) such that \(A_{2,ii} = A_{2,jj}\), \emph{i.e.}, species \(i\) and \(j\) have the same bare velocity \(\beta D_iq_i = \beta D_j q_j\).
If yes, we refer to it as degenerate case; if not we call it a non-degenerate case.
We will start with the non-degenerate case as it is simpler.
\subsection{Non-degenerate case}
\label{sec:orgede3d64}
If \(A_{2,ii} \neq A_{2,jj}\) for any \(i \neq j\),
then
\(\tilde{C}^{(0)}_{ij} = 0\), \emph{i.e.}, \(\tilde{C}^{(0)}\) must be diagonal.
At the first order of \(\varepsilon\), we have
\begin{align}
\label{eq:order-1}
A_1 \tilde{C}^{(0)} + \tilde{C}^{(0)} A_1^{\text{T}} + A_2 \tilde{C}^{(1)} - \tilde{C}^{(1)} A_2 = 2 R.
\end{align}
To take advantage of \(A_2\) being diagonal, we consider the diagonal elements of the above equation. This then leads to
\(\left[A_2 \tilde{C}^{(1)} - \tilde{C}^{(1)} A_2 \right]_{ii} = 0\)
and
\(A_{1,ii}\tilde{C}^{(0)}_{ii} = R_{ii}\).
Therefore,
\begin{align}
\tilde{C}^{(0)}_{ij} = \frac{R_{ii}}{A_{1,ii}} \delta_{ij}
= \frac{\bar{\rho}_i(k^2 + p_n^2)}{k^2 + p_n^2 + \kappa_i^2} \delta_{ij},
\end{align}
with \(\kappa_i^2 \equiv \frac{\beta q_i^2 \bar{\rho}_i}{\epsilon}\) being the inverse square of Debye length for species \(i\).
Consequently, we obtain
\begin{align}
S_n^{(0)}
= \frac{1}{2} \sum_{i=1}^N \left(\frac{k^2}{k^2 + p_n^2} - \frac{k^2 + \kappa_i^2}{k^2 + p_n^2 + \kappa_i^2}  \right),
\end{align}
where we have discarded terms that are irrelevant for the van der Waals interaction.
Therefore, the total van der Waals interaction in the large \(E\) limit in non-degenerate case is
\begin{align}
f_{\text{t}} =&\, \frac{\zeta(3)}{8\pi\beta H^3} (N-1) \nonumber\\
&\, -
\sum_{i=1}^N\frac{k_{\text{B}}T}{4\pi H^3} \int_{H\kappa_i}^{\infty} \mathrm{d} K K^2 (\coth K -1),
\end{align}
from which one can see each species contributes a long-range repulsive force with a magnitude the same as that of van der Waals force without electrolytes and a screened attractive force dependent on individual Debye lengths.
\subsection{Degenerate case}
\label{sec:orgad057e5}
Assume there exists a pair of species indices \(i\) and \(j\) (\(i\neq j\)) such that \(A_{2,ii} = A_{2,jj}\), \emph{i.e.}, species \(i\) and \(j\) have the same bare velocity \(\beta D_i q_i =\beta  D_j q_j\).
Now the zeroth order equation, Eq.~\eqref{eq:zero-order},
does not guarantee that \(\tilde{C}_{ij}^{(0)}\) vanishes anymore.
The \((i,j)\) and \((j,i)\) element of the second term in the square bracket
of the first order equation, Eq.~\eqref{eq:order-1}, vanishes
\begin{subequations}
\begin{align}
\left[A_2 \tilde{C}^{(1)} - \tilde{C}^{(1)} A_2 \right]_{ij}
&= \left( A_{2,ii} - A_{2,jj}  \right) \tilde{C}^{(1)}_{ij} = 0, \\
\left[A_2 \tilde{C}^{(1)} - \tilde{C}^{(1)} A_2 \right]_{ji}
&= \left( A_{2,jj} - A_{2,ii}  \right) \tilde{C}^{(1)}_{ji} = 0.
\end{align}
\end{subequations}
Following the same reasoning as in the non-degenerate case, for \((k,l)\notin \{(i,j),\,(j,i)\}\)
\begin{align}
\tilde{C}^{(0)}_{kl} = \frac{R_{kl}}{A_{1,kl}} \delta_{kl},
\end{align}
that is the degenerate subspace consisting of species \(i\) and \(j\) decouples from the rest.

We now consider a general degenerate subspace of dimension \(M\), \(2 \leq M \leq N-1\), \emph{i.e.},
there is a set of species index \(I\) of maximal cardinality \(M\) such that \(\forall i, j\in I\), \(A_{2,ii} = A_{2,jj}\).
Without loss of generality, we relabel these indices by \(I=\{1,2,\ldots,M\}\).
We solve the equation
\begin{align}
\label{eq:deg-M}
A'_1 \tilde{C}^{(0)} + \tilde{C}^{(0)} A_1^{'\text{T}} = 2 R',
\end{align}
where the primed matrices \(A'_1\) and \(R'\) are, respectively, the principal sub-matrices of \(A_1\) and \(R\) with indices confined in \(I\).
Since \(A_1\) has to be positive definite to ensure a well-defined dynamical equations,
\(A'_1\) as the principal sub-matrices of \(A_1\) should also be positive definite.
Therefore, the situation here reduces exactly to that of the equilibrium case.

It follows that in the large \(E\) limit, the ions belonging to different subspaces are decoupled
while those within the same subspace are correlated with the equilibrium correlation function of the system which only contains those species. 
We denote the total number of the degenerate and non-degenerate subspaces by \(\tilde{N}\) and the species indices set for \(\alpha\)-subspace by \(I_\alpha\) and also define \(\kappa_{\alpha}^2 = \sum_{i \in I_{\alpha}} \kappa_i^2\).
In this notation, the total van der Waals interaction in the large \(E\) limit in degenerate case is
\begin{align} \label{eq:Casimir-degenerate}
f_{\text{t}} =&\, \frac{\zeta(3)}{8\pi\beta H^3} (\tilde{N}-1) \nonumber\\
&\, -
\sum_{\alpha=1}^{\tilde{N}}\frac{k_{\text{B}}T}{4\pi H^3} \int_{H\kappa_{\alpha}}^{\infty} \mathrm{d} K K^2 (\coth K -1),
\end{align}
from which one can see each \emph{species subspace} sharing the same bare velocity contributes a long-range repulsive force with a magnitude the same as that of van der Waals force without electrolytes and a screened attractive force dependent on corresponding Debye lengths.

Based on this idea of decoupling  of ions at large \(E\) limit, an effective equilibrium-like partition function can be used to describe the system:
\begin{align}
&\,Z_{\text{ion}} = \int \mathcal{D}[n_i] \exp \Bigg\{
-\frac{1}{2} \int \mathrm{d}\mathbf{x} \sum_{i=1}^N \frac{[n_i(\mathbf{x})]^2}{\bar{\rho}_i} \nonumber\\
&\, + \frac{\beta}{2} \sum_{\alpha=1}^{\tilde{N}} \sum_{i,j\in I_{\alpha}} \int \mathrm{d}\mathbf{x}\mathrm{d}\mathbf{x}' q_i n_i(\mathbf{x})G(\mathbf{x},\mathbf{x}') q_j n_j(\mathbf{x}')
\Bigg\},
\end{align}
where only ions with the same bare velocity interact with each other. Computing the corresponding effective {\it free energy} plus the free energy from the dipole degrees of freedom,  one finds  the total van der Waals interaction given by the dynamical computation leading to 
Eq.~\eqref{eq:Casimir-degenerate}.
\section{Long-Range van der Waals interaction with external electric fields of general strength}
\label{general-E}
For external electric fields of general strength, one has to explicitly solve the
the Lyapunov equation, Eq.~\eqref{eq:Lyapunov}, to obtain the correlations in the steady state.
As an example, we consider the binary asymmetric electrolyte case where $\overline \rho_1 q_1+\overline \rho_2 q_2 =0 $ and $D_1\neq D_2$.
In this case the  correlation  functions in the steady state take a very  complicated algebraic form . However, as we are interested in the long-range behavior of the interaction and this can be deduced by carrying out an expansion for small  wave vector $k$ 
of the van der Waals interaction (to see this more clearly one can make the change of variables ${\bf k}={\bf p}/H$ in the integral and sums involved), this then yields the simple result
\begin{equation}
 f^{\text{lr}}_{\text{t}}(H) =
\frac{\zeta(3)}{8\pi \beta H^3} \left[ 1 - \frac{1}{\sqrt{1+\alpha^2 } }\right],
\end{equation}
where
\begin{equation}
\alpha^2 = \frac{\kappa_1^2\kappa^2_2 (\Delta v)^2}{(\kappa_1^2+\kappa_2^2)(D_1\kappa_1^2+ D_2\kappa_2^2)^2},
\end{equation}
 and 
 \begin{equation}
 \Delta v = v_1-v_2
 \end{equation}
 is the difference in the bare velocities of specie 1 and species 2. One can slightly generalize the model considered here by introducing additional external forces on the ionic species in the direction of the field denoted by $F_1$ and $F_2$ (for instance the buoyancy  force acting on colloids), in addition to the force due to the applied electric field. In this case one finds 
\begin{equation}
\Delta v = v_1-v_2= \beta [D_1 (q_1 E+ F_1) -D_2 (q_2 E+F_2)].
\end{equation}
 However in the case where $F_1=F_2=0$, the electroneutrality condition leads to the remarkable result
\begin{align}
f^{\text{lr}}_{\text{t}}(H) =
\frac{\zeta(3)}{8\pi \beta H^3} \left[ 1 - \frac{1}{\sqrt{1+ \frac{\beta \epsilon E^2}{\bar{\rho}_1 + \bar{\rho}_2}}} \right],\label{2elec}
\end{align}
which we see is independent of the diffusion constants of the ions. We also notice that the magnitude of the deviation from the equilibrium result is determined by the parameter 
\begin{equation}
    \alpha^2 = \frac{\epsilon E^2}{(\bar{\rho}_1+\bar{\rho}_2) k_B T},
\end{equation}
which is proportional to the ratio of the Maxwell pressure $P_{\rm Max} = \epsilon E^2/2$ and the ideal gas pressure $P_{\rm id}=(\bar{\rho}_1+\bar{\rho}_2) k_B T $. This feature, however, holds only in binary electrolyte case. When the number of species becomes larger \(N \geq 3\), the corresponding long-range van der Waals interactions have complicated dependence on the charges \(q_i\), mean densities \(\bar{\rho}_i\), and diffusion constants \(D_i\).

We now  give some special cases when \(N=3\)  where simple tractable formulas can be obtained.
Consider the case where  mean densities and diffusion constants of the three species are equal, \emph{i.e.}, \(\bar{\rho}_1 = \bar{\rho}_2 = \bar{\rho}_3 = \bar{\rho}\), \(D_1 = D_2 = D_3\).
When \(q_1 = q_2 = 1\) and \(q_3 = -2\), the system is clearly electroneutral. Apart from being discernible the ions  of type $1$ and $2$ are otherwise electrically and transport wise identical, the system is up  to this indiscernibility essentially the same case studied above for the case $q_2=-2q_1$. Here we find
\begin{align}
f^{\text{lr}}_{\text{t}}(H) =
\frac{\zeta(3)}{8\pi \beta H^3} \left[ 1 - \frac{1}{\sqrt{1+ \frac{\beta \epsilon E^2}{3\bar{\rho}}}} \right].
\end{align}
Notice that this is the same as the result in Eq. (\ref{2elec}) for a binary system with $q_2=-2q_1$, $\bar\rho_1=2\bar\rho$ and $\bar\rho_2=\bar\rho$ and
 agrees with the general result in Eq.~\eqref{eq:Casimir-degenerate}
 as the species 1 and 2 have the same bare velocity.

When \(q_1 =1\), \(q_2 = 2\) and \(q_3 = -3\) one again finds a simple result:
\begin{align}
f^{\text{lr}}_{\text{t}}(H) =
\frac{\zeta(3)}{8\pi \beta H^3} \left[ 2 - \frac{1}{\sqrt{1+ \frac{\beta \epsilon E^2}{3\bar{\rho}}}} - \frac{1}{\sqrt{1+ \frac{50 \beta \epsilon E^2}{1029 \bar{\rho}}}} \right].\label{elec3}
\end{align}
and in the limit of large \(E\), the result again agrees with the general result in Eq.~\eqref{eq:Casimir-degenerate}.  

One further point about diffusion constants worth mentioning is that the Lyapunov equation, Eq.~\eqref{eq:Lyapunov}, is homogeneous with respect to \(D_i\). Therefore, the correlations in the steady state and hence the van der Waals interaction depend only on the ratios, say, \(D_i/D_1\), \(i=2,3,\ldots N\). 

Despite the fact that the system can be effectively described in terms of linear Lyapunov equations, the extension to the general case of $N>2$ ionic types at finite applied  fields $E$ turns out to be highly complex, and the resulting formulas are too complicated and long to analyze and even write down outside of computer algebraic packages.

\section{Conclusions and discussions}
\label{conclusions}
To sum up, we consider multi-species electrolytes driven out of equilibrium by applied electric fields.
By using the stochastic density functional theory, we first of all reproduce the equilibrium results, verifying the validity of the method.
Then we find that in the large applied electric field limit, the correlation between species with different bare velocities are decoupled,
but the correlation between species sharing the same bare velocity  remains the same as that when only those species are present and in equilibrium.
This corroborates that large applied electric fields suppress Debye screening between ions with different bare velocities.
The resulting long-range repulsive van der Waals interaction in the $E \to \infty$ limit is thus linearly proportional to the number of different bare velocities,
providing an additional way to tune the van der Waals interaction in multi-species electrolytes driven by applied electrical fields.
As a concrete example, consider two solutions, one of NaCl with concentration $\bar{\rho}$ and the other of NaCl, with concentration $\bar{\rho}-\bar{\rho}'$, along with NaBr, with concentration $\bar{\rho}'$.
In the first solution when $E\to\infty$, the long-range van der Waals force is $k_\text{B} T \zeta(3)/8\pi H^3$, while in the second solution it is $2\times k_\text{B} T \zeta(3)/8\pi H^3$. This difference emerges solely from the fact that Cl\(^-\) and Br\(^-\) have different diffusion constants and consequently different bare velocities.
Further extensions of current work may include taking into account the double layer inhomogeneity, the finite size of ion particles \cite{avniConductivityConcentratedElectrolytes2022}, the effect of a  polar solvent \cite{illienStochasticDensityFunctional2024} as well as hydrodynamic effects. Although the model studied here is rather simplified and idealized, the basic idea of charge decoupling seems to be quite robust and could provide a useful framework to understand more complicated problems involving electrically driven electrolytes.

\begin{acknowledgments}
It is with much regret and sadness that  G.D., D.S.D.~and B.M.~inform the reader that our coauthor and friend Professor Rudolf Podgornik passed away on 28th December 2024, shortly before the final version of this manuscript was finalized. G.D.~and  B.M.~acknowledge funding from the Key Project No.~12034019 of the National Natural Science Foundation of China and the support by the Fundamental Research Funds for the Central Universities (Grant No.~E2EG0204). D.S.D.~acknowledges support  from the grant No.~ANR-23-CE30-0020 EDIPS, and by the European Union through the European Research Council by the EMet-Brown (ERC-CoG-101039103) grant.
\end{acknowledgments}

\bibliography{multiple_species_refs}
\end{document}